\documentclass[aip,jap,reprint,10pt,showpacs,altaffillsymbol,superscriptaddress,longbibliography]{revtex4-1}
\pdfoutput=1
\usepackage{graphicx}
\usepackage{amsfonts}
\usepackage{amsmath}
\usepackage{natbib}
\usepackage{color}

\begin{document}

\title{Precision measurement of an electron pump at 2 GHz}

\author{Stephen~P.~Giblin}
\affiliation{National Physical Laboratory, Hampton Road, Teddington, Middlesex TW11 0LW, United Kingdom}
\author{Gento Yamahata}
\affiliation{NTT Basic Research Laboratories, NTT Corporation, 3-1 Morinosato Wakamiya, Atsugi, Kanagawa 243-0198, Japan}
\author{Akira Fujiwara}
\affiliation{NTT Basic Research Laboratories, NTT Corporation, 3-1 Morinosato Wakamiya, Atsugi, Kanagawa 243-0198, Japan}
\author{Masaya Kataoka}
\affiliation{National Physical Laboratory, Hampton Road, Teddington, Middlesex TW11 0LW, United Kingdom}

\email[stephen.giblin@npl.co.uk]{Your e-mail address}

\date{\today}

\begin{abstract}
A well-characterised sample of silicon tunable-barrier electron pump has been operated at a frequency of $2$~GHz using a custom drive waveform, generating a pump current of $320$~pA. Precision measurements of the current were made as a function of pump control parameters, using a blind protocol, over a 7-week campaign. The combined standard uncertainty for each $\sim 10$~hour measurement was $\sim 0.1$~parts per million. The pump current exhibits a plateau along the exit gate voltage flat to approximately $0.1$~parts per million, but offset from $ef$ by $0.2$~parts per million. This offset may be a sign of errors in the current traceability chain, indicating a limit to the accuracy of small current scaling using existing methods based on cryogenic current comparators.
\end{abstract}

\pacs{1234}

\maketitle

\section{Introduction}

Electron pumps are devices that aim to generate a reference DC electric current by moving electrons one at a time in response to a periodic control signal at frequency $f$. They potentially offer a simple and elegant traceability route for small currents, direct to the SI definition of the ampere\cite{kaneko2016review,scherer2019single}. A class of pumps fabricated from semiconductor materials\cite{kaestner2015non} has demonstrated accurate and robust current generation at roughly the part-per-million (ppm) accuracy level, for currents $I_{\text{P}}$ up to $160$~pA \cite{giblin2019evidence}. However, important questions must be answered before electron pumps can confidently be adopted as reference current standards at the uncertainty levels of primary electrical metrology. Most significantly, the robustness and device independence of the current needs to be demonstrated at at least the $0.1$~ppm level, over a range of device designs and operating parameters. To date, two studies have focused on the robustness of the current from GaAs pumps, at current levels of $\sim 100$~pA, at uncertainty levels for each data point of $\sim 2$~ppm\cite{giblin2017robust} and $\sim 0.5$~ppm\cite{stein2016robustness}. 

However, blind measurement techniques which have been implemented in other metrology areas to remove bias \cite{schlamminger2015summary} have not yet been applied to the study of electron pumps where the pump current is treated as an unknown and compared to a known reference current. Addressing unconscious experimenter bias is particularly important in experiments where the expectation of the result is strongly constrained; in this case, we expect $I_{\text{P}} = ef$, and there is a possibility that in a non-blind measurement, the experimenter may unconsciously favour pump control parameters that yield this result. Particularly important in the electron pump context is the lack of reproducibility in attempts to realise a capacitance standard based on pumping a known number of electrons onto a cryogenic capacitor \cite{keller1999capacitance,scherer2017electron}. The authors of Ref. \onlinecite{scherer2017electron} were unable to reproduce the results of Ref. \onlinecite{keller1999capacitance}, and identified components in the capacitance measurement uncertainty which had previously been under-estimated.

Evaluating the robustness of the pump current presents a challenge due to the time-scales involved: the small currents require many hours of averaging to resolve $0.1$~ppm for a single data point, and the time-scale of the whole measurement campaign challenges the stability of the measurement system and the electron pump itself\cite{giblin2019evidence}. To reduce the measurement time, or equivalently, to allow more data points to be measured within the timescale of a measurement campaign, the pump current should be increased as much as possible. Custom gate drive waveforms which slow down the electron capture process have been used to operate GaAs pumps accurately at much higher frequencies than were possible with sine wave drive \cite{giblin2012towards,stein2015validation,stein2016robustness}. With these pumps the upper frequency limit for accurate pumping was $f \sim 1$~GHz even with the custom waveforms. Silicon pumps, on the other hand, have demonstrated accurate pumping at $f=1$~GHz with sine wave drive \cite{zhao2017thermal,giblin2020realisation}, and the possibility of increasing the frequency further while maintaining sub-ppm pumping accuracy using custom waveforms has not yet been explored. 

\begin{figure}
\includegraphics[width=9cm]{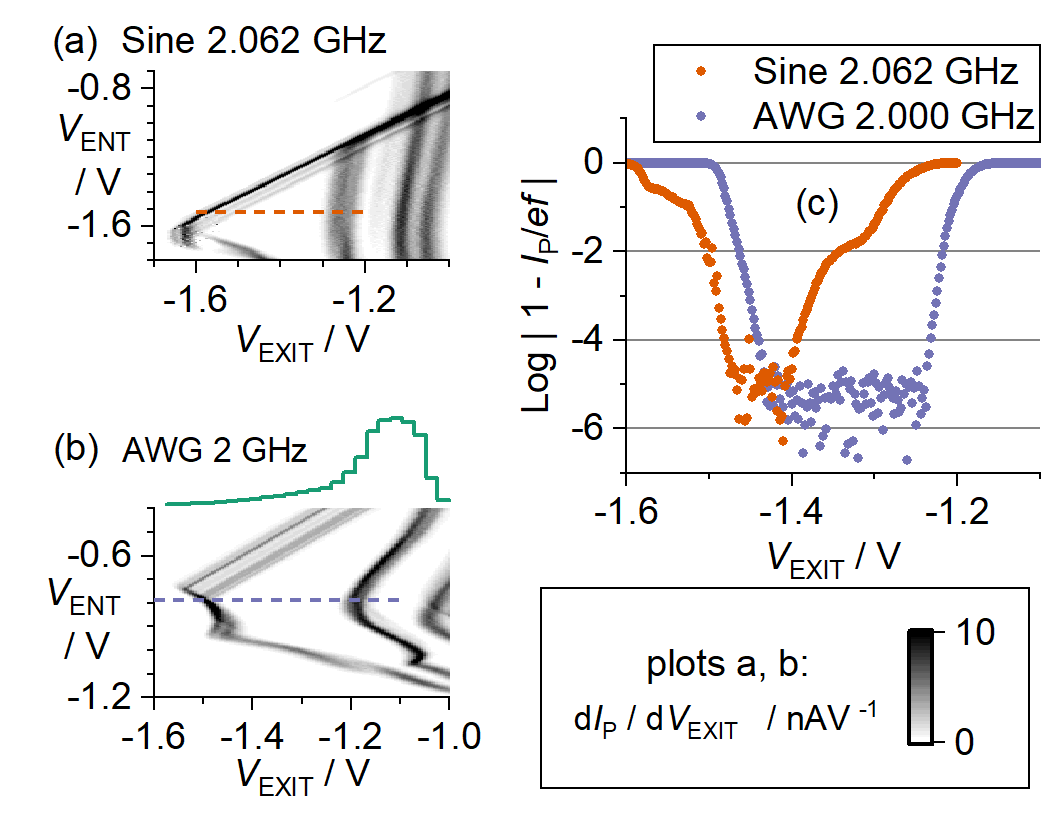}
\caption{\label{AWGGraphFig}\textsf{(a): Grey-scale derivative pump map using sine wave drive at 2.062 GHz, $P_{\text{RF}}=13.2$~dBm. (b): Pump map using a waveform from an AWG at repetition rate $f=2$~GHz. One cycle of the AWG waveform is shown in the inset. Note that this waveform is subsequently amplified by an inverting amplifier to yield the correct polarity of gate voltage, whereby the negative voltage pulse on the entrance gate raises the entrance barrier to pump an electron. (c): Log-scale plots of the pump current along the horizontal dashed lines in plots (a) and (b).}}
\end{figure}

\section{Experimental method and blind protocol}

We investigate a single well-characterised sample of silicon pump which has previously been the subject of two precision measurement campaigns\cite{yamahata2016gigahertz,giblin2020realisation}. The pump is a silicon nanowire-MOSFET, in which charge carriers are induced by a positive voltage applied to a global top gate \cite{fujiwara2008nanoampere,yamahata2016gigahertz} which was set to $4$~V for all the measurements. Two finger gates, denoted the entrance gate and exit gate, define the region of the nanowire where a single electron can be trapped. Negative DC voltages $V_{\text{ENT}}$ and $V_{\text{EXIT}}$ applied to these gates define the pump operating point, and the periodic pump drive signal is added to $V_{\text{ENT}}$ using a room-temperature bias-tee. A $50$~Giga samples/s arbitrary waveform generator (AWG, Tektronix 70001A) was used to generate a custom waveform for the pump drive. Because the AWG output had a maximum peak-peak amplitude of $V_{\text{AC}} = 0.5$~V, the output was amplified by a wide-band inverting RF amplifier with $+15$~dB gain before the bias-tee. The AWG is referenced to a $10$~MHz frequency reference derived from a hydrogen maser.

Figure \ref{AWGGraphFig} shows characterisation data using both sine wave drive and the custom AWG waveform at a repetition frequency of $2$~GHz. It is clear from the log-scale plots of figure 1 (c) that there is a substantial plateau along the exit gate axis when using the AWG drive waveform, but not when using sine wave drive. The inset to figure 1(b) shows the AWG waveform used for all the measurements reported in this paper. Characterisation data at other frequencies is inlcuded in supplementary sections A and B.

The experimental apparatus and methods used for this study are in many respects identical to that used in Ref. \onlinecite{giblin2020realisation}. As in those experiments, the pump is cooled to a temperature close to 4 K by suspending it above a liquid helium surface. The pump current $I_{\text{P}}$ is measured using a noise-optimised ultrastable low-noise current amplifier (ULCA) \cite{krause2019noise}, with a precision digital voltmeter (DVM) recording the ULCA output. As in Ref. \onlinecite{giblin2020realisation}, the DVM was calibrated roughly once every hour by switching its input to a Josephson voltage standard (JVS). A single precision measurement typically lasted between 8 and 10 hours and included between 7 and 11 voltmeter calibrations. To remove offset drifts in the measurement system during precision measurements, the pump drive signal was toggled on and off with a cycle time of $228$~s. Roughly the first $34$ seconds of each data segment (300 out of 1000 data points) following each on or off switch was rejected from the analysis to remove transient effects. More details of the measurement protocol are given in supplementary section C. 

The pump current $I_{\text{P}}$ is calculated from the on-off difference in the DVM voltages $\Delta V$ using the equation $I_{\text{P}} = \Delta V / A_{\text{TR}}$. Here, $A_{\text{TR}}$ is the trans-resistance gain of the ULCA, nominally equal to $10^9$~V/A. This gain is calibrated against the quantum Hall resistance (QHR) in 2 stages\cite{drung2015ultrastable} and via some intermediate transfer standards, using a cryogenic current comparator (CCC) with relative uncertainty less than $0.1$~ppm \cite{giblin2019interlaboratory}. Detailed calibration results are reported in supplementary section E. The measurement of the pump current was therefore traceable to the SI unit ampere via the JVS, the QHR, and the relationship $I=V/R$. For characterisation measurements such as those reported in figures 1, 2a, and small filled points in figures 2b and 2c, no offset subtraction was performed: the pump drive signal was left on, and each data point is a single $20$ power line cycle DVM measurement. 

A blind protocol was implemented so that the lead experimenter could not see the true value of $I_{\text{P}}$ while the measurements and data analysis were in progress. This is achieved by multiplying all the DVM readings by a hidden scaling factor $\beta=1.00000387$, at a low level in the measurement software. An exception occurs when when the DVM is connected to the JVS for calibration, in which case $\beta =1$. While tuning the pump and performing measurements, the experimenter does not know the scaling factor and can only access the scaled pump current $I_{\text{P,B}} = \beta \Delta V / A_{\text{TR}}$. Therefore, the tuning of the pump operating parameters and the choice of parameters for the precision measurements can only be made with reference to the flatness of the current plateau, not the deviation of the current from $ef$. The scaling factor was programmed and password-protected by a member of the team who was not otherwise involved in the experiments. The experimenter knew that it was constrained such that $|1 - \beta| < 5 \times 10^{-6}$ so that gross failures of the pump or apparatus would be apparent during characterization measurements. The scaling factor was revealed after the experiments were finished and data analysis, including analysis of the ULCA calibrations, completed.

\section{precision measurement campaign}

The aim of the measurements was to study the pump current as a function of control parameters $V_{\text{ENT}}$, $V_{\text{EXIT}}$ and $V_{\text{AC}}$. To this end, a total of 67 precision measurements were made during a 7-week campaign, employing the apparatus and blind protocol described in section II. The measurements were divided into 17 `runs'. For most of the runs, several measurements were made while varying one control parameter. Runs 11-13 consisted of single measurements without varying a parameter. Further detail of the measurement chronology is given in supplementary section D. To monitor the stability of the pump, a `fingerprint' pump map was obtained before and after each run, apart from a few occasions when it was prevented by an experimental difficulty. For completeness, all of these pump maps are shown in supplementary section H. Additional line scans of current as a function of one or more control parameters were also measured to assess the optimal values of fixed control parameters for the next precision run. Typically, these scans were used to find the value of the control parameter that maximised the plateau width. They used a single $20$~PLC measurement for each data point, with a relative uncertainty of approximately $10$~ppm per data point. A pass / fail stationary mean statistical test, described in supplementary section G, was applied at the data analysis stage to each precision measurement to evaluate whether the current was stable during the measurement time. 

\begin{figure}
\includegraphics[width=9cm]{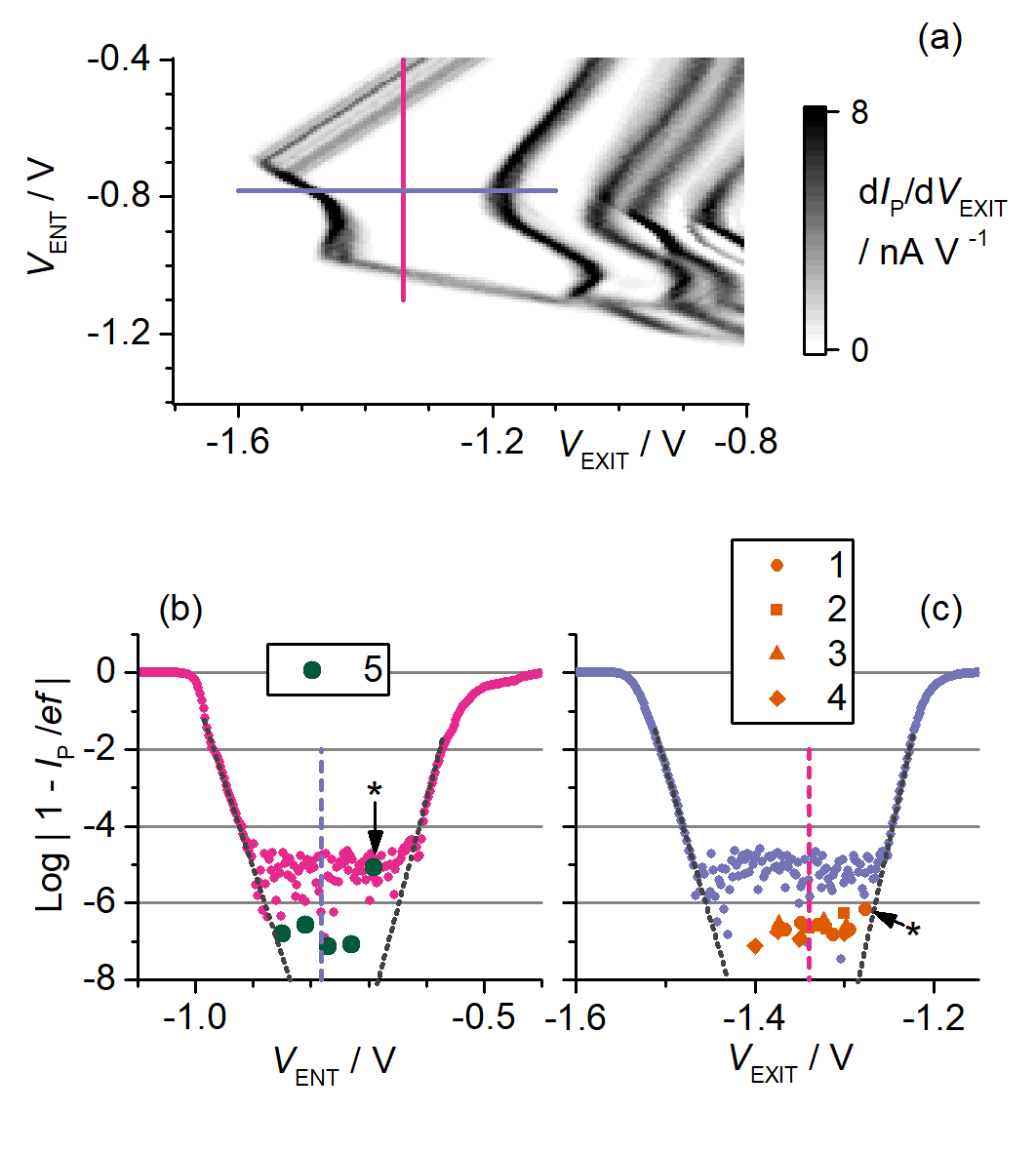}
\caption{\label{LogPlotFig}\textsf{(a): Derivative pump map measured after precision run 4 and before precision run 5. (b) and (c): Line-scans of the pump current measured along the gate voltage axes indicated by solid colored lines in (a), and plotted on a logarithmic scale. The vertical dashed lines indicate the fixed value of entrance (exit) gate voltage used for the exit (entrance) gate scan. The diagonal dashed lines are guides to the eye extrapolating the exponential edges of the plateau. Larger filled points are the precision measurement data for runs 1-5, with the run number indicated in the plot legend. Data points indicated with a star ($*$) failed the stationary-mean test.}}
\end{figure}

\section{Results of precision measurements}

\subsection{Precision results}

After run 5, the pump became less stable, (supplementary section H), making it difficult to establish the flatness of plateaus along $V_{\text{ENT}}$ and $V_{\text{EXIT}}$ axes. For this reason, we concentrate here on the data from runs 1-5, although the full precision data set is presented in supplementary figure S9. In figure \ref{LogPlotFig}, we present the data from the first 5 precision runs. Panel (a) shows a pump map recorded between runs 4 and 5, and panels (b) and (c) show line-scans on a log scale which highlight the deviation of the current from the ideal value on the $1ef$ plateau. The fixed value of $V_{\text{ENT}}$ ($V_{\text{EXIT}}$) for the $V_{\text{EXIT}}$ ($V_{\text{ENT}}$) line-scan was adjusted in order to maximise the width of the plateau in the log-scale plot. The results of precision runs 1-5 are plotted as solid points in figures \ref{LogPlotFig} (b) and (c). Runs 1-4 were $V_{\text{EXIT}}$ scans, plotted in figure \ref{LogPlotFig} (c), and run 5 was a $V_{\text{ENT}}$ scan, plotted in figure \ref{LogPlotFig} (b). The 18 data points along the $V_{\text{EXIT}}$ axis (figure \ref{LogPlotFig} (c)) define a plateau in agreement with an extrapolation of the standard-accuracy measurement. The precision data point marked with a star ($*$), failed the stationary-mean test, presumably because it was close to the edge of the plateau, and small fluctuations in offset charge, equivalent to shifts in $V_{\text{EXIT}}$, caused fluctuations in the pumped current to be resolved on the time-scale of the precision measurement.

The precision $I_{\text{P}}(V_{\text{EXIT}})$ data for runs 1-4 (apart from the point that failed the stationary mean test) are re-plotted on a linear y-axis in figure 3a as $\Delta I_{\text{P}} = (I_{\text{P}} - ef)/ef$. The mean of these 17 points is $\Delta I_{\text{P}} = 0.22$~ppm, with a standard deviation $\sigma$ of $0.14$~ppm. The individual data points have a mean combined uncertainty $\langle U_{\text~{T}}\rangle$ of $0.102$~ppm, although the uncorrelated random uncertainty, $U_{\text{A}}$, for each data point is smaller, in the range $0.08-0.09$~ppm. The scatter of the points is therefore slightly larger than what would be expected from the type A uncertainty of each point ($\sigma > \langle U_{\text~{A}}\rangle$), although not statistically incompatible with the assumption that the data is sampling a stationary mean along the plateau. We can therefore conclude that this data is consistent with a plateau along the $V_{\text{EXIT}}$ axis, flat at the $0.1$~ppm level, but significantly offset from $ef$ by $0.22$~ppm. Figure 4 (b) shows the same data re-analysed with the first 700 data points rejected from the beginning of each 1000-point data segment instead of the standard 300. This was to test for the presence of a time constant in the current, as discussed in section V.

\begin{figure}
\includegraphics[width=9cm]{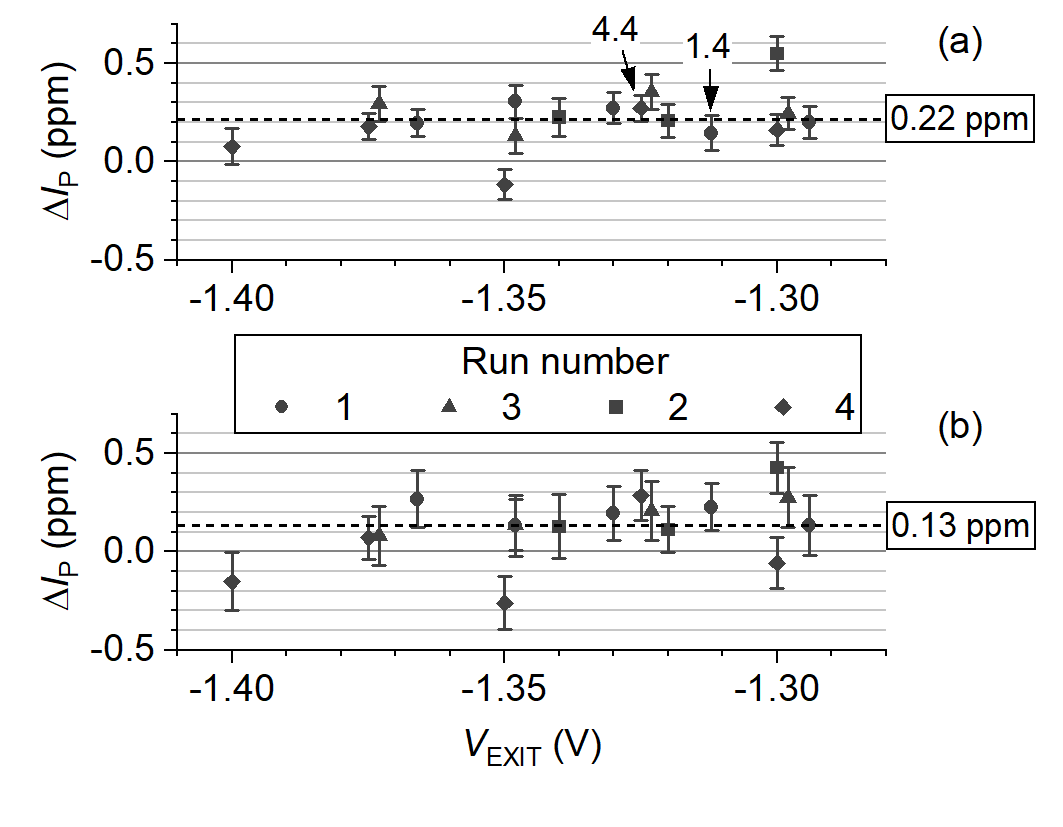}
\caption{\label{HighAccFig}\textsf{Results of precision measurements for runs 1-4 as a function of $V_{\text{EXIT}}$, expressed as $\Delta I_{\text{P}} = (I_{\text{P}} - ef)/ef$. The data have been analysed with (a): 300 and (b): 700 data points rejected from the start of each 1000-point data segment. Error bars show the combined standard uncertainty $U_{\text{T}}$ The horizontal dashed lines show the weighted means of each data set. Arrows highlight run 1, measurement 4 and run 4, measurement 4. A breakdown of the uncertainty for these measurements is given in table I.}}
\end{figure}

Only one precision run was performed along the $V_{\text{ENT}}$ axis before the interruption, illustrated by the heavy filled points in figure \ref{LogPlotFig} (b). One data point, marked with a $*$, failed the stationary-mean test. It is not clear from this single run whether this data point indicates real structure to the plateau at level of $\sim 5$~ppm, or if it is the result of a drift in the device state. The remaining 4 data points mark a plateau region which, combined with the stability of the pump map from runs 1-5, gives confidence that the fixed value of of $V_{\text{ENT}}$ selected for runs 1-4 is in the middle of an experimentally-determined plateau. The mean of the 4 measurements from run 5 is $\Delta I_{\text{P}} = 0.15$~ppm, with a standard deviation of $0.09$~ppm. This is consistent with the deviation measured in runs 1-4 given the much smaller sample size.

\subsection{Uncertainty}

In table I, the breakdown of the uncertainty is given for two measurements indicated by arrows in figure 3a. The uncertainties due to the two stages of the ULCA calibration are presented as separate components, with the uncertainty due to the drift of the ULCA gains in between calibrations included in these two terms. This was significantly reduced by performing frequent ULCA calibrations, with more detail given in supplementary sections D and E. Run 1, measurement 4 is a typical representative measurement, and run 4, measurement 4 had the lowest combined uncertainty of the campaign. As in previous measurement campaigns, the type A uncertainty of the pump measurement is the largest single contribution, but the larger pump current achieved in this study has reduced $U_{\text{A}}$ to below $0.1$~ppm and the uncertainty in the ULCA calibration is now a significant contribution. Specifically, the uncertainty in the output stage gain $R_{\text{IV}}$ (nominal value $1$~M$\Omega$) is limited by the $0.04$~ppm uncertainty in the $100$~k$\Omega$ reference resistor traceable to the QHR via a chain of 3 intermediate measurements \cite{giblin2018reevaluation,giblin2019interlaboratory}.

\begin{table}
\caption{\label{UncertTable} Uncertainty breakdown for run 1, measurement 4, and run 4, measurement 4. All entries in the table are dimensionless relative uncertainties ($k=1$) in parts per million.}
\centering
\setlength{\tabcolsep}{8pt}
\begin{tabular}{c c c}
Contribution & Meas. 1.4 & Meas. 4.4 \\
\hline\hline
ULCA $G_{\text{I}}$ Cal. & 0.024 & 0.024 \\ 
ULCA $R_{\text{IV}}$ Cal. & 0.062 & 0.043 \\ 
ULCA Temp. corr. & 0.023 & 0.023 \\ 
DVM Cal. & 0.014 & 0.016 \\
Pump $U_{\text{A}}$  & 0.088 & 0.061 \\
\hline
Total $U_{\text{T}}$& 0.111 & 0.084 \\
\end{tabular}
\end{table}

\subsection{stability of the pump}

The measurement campaign was divided into two parts by an instrument issue which forced a period of 8 days' down-time between runs 5 and 6. During this time, the pump was thermally cycled to room temperature and back to $4$~K twice. From examination of the pump maps in supplementary section H, it is clear that the pump became less stable after this interruption, although even before the interruption, small changes in the `nose' (the onset of pumped current as $V_{\text{EXIT}}$ is made less negative) of the pump map are visible. This contrasts with the data of Ref. \onlinecite{giblin2020realisation} showing this sample of pump to be extremely stable over multiple cool-downs in different laboratories, when driven with a sine wave at $\sim 1$~GHz. We conjecture that at least some of the changes visible in the pump maps during the present campaign may be due to changes in the transmission of the cryogenic microwave line at frequencies $\gg 1$~GHz. This could plausibly arise due to changes in the temperature gradient along the line, and would affect the high frequency components of the AWG waveform, causing distortion of the waveform at the pump entrance gate.

\section{discussion}

The study was complicated by instability in the pump map, which made it difficult in the later parts of the measurement campaign to interpret the results as sampling a stable state of the device. However, enough results were obtained from runs 1-5 to establish that the pump current is invariant in the exit gate voltage at the level of $1$ part in $10^{7}$. Averages over these data points presented in the previous section give $\Delta I_{\text{P}} \sim 2 \times 10^{-7}$, a significant offset from the ideal current $I_{\text{P}} = ef$. The flatness of the plateau suggests that the offset is due to an error in the measurement system which applies a constant offset to all the measurements, rather than an error due to the physics of the pump itself.

One possible cause of error is a time constant in the pump current. This was discussed in Ref. \onlinecite{giblin2020realisation}, and could plausibly arise from heating due to the relatively large RF powers applied to the device gate. Repeating the data analysis of runs 1-4 with 700 data points rejected from the start of each segment instead of 300 did indeed yield an average pump current closer to $ef$, as shown in figure 3b. However, the larger type A uncertainties in this analysis make it difficult to draw a firm conclusion regarding possible time constants. Measurements with much longer on-off cycle times could potentially resolve this question, but require the $1/f$ noise corner of the ULCA current measurement to be at frequencies well below $1$~mHz. ULCA units have demonstrated this performance in bench tests \cite{krause2019noise}, but the cryogenic wiring involved in a pump measurement introduces additional sources of noise and drift. Another possible cause of error is a non-linearity in the gain of the ULCA. The ULCA input stage gain $G_{\text{I}}$ is calibrated at a current of $6$~nA, and the pump current is $320$~pA. Comparisons of the gains of two ULCA units with different input stage designs, detailed in supplementary section F, set an upper limit to possible non-linearity of a few parts in $10^8$, so this is unlikely to cause errors of a part in $10^{7}$. 

Possibly the most important cause of error could arise from the CCC calibration of the ULCA $G_{\text{I}}$. This could result from rectification of noise by the CCC's SQUID detector leading to different SQUID offsets for the two polarities of current used in the ULCA calibration \cite{drung2014ultrastable}. One study on CCC performance in the low-flux regime\cite{drung2015improving} concluded that noise pickup might lead to this type of error at SQUID flux levels below $1$~$\mu \Phi_{0}$, although this number was based on a limited number of measurements and is specific to a particular CCC design\cite{goetz2014compact}, different in detail to the CCC used to calibrate the ULCA in our experiments. We calibrated the ULCA $G_{\text{I}}$ using a CCC\cite{giblin2019interlaboratory} with a $10000:10$ turns ratio, and a sensitivity of $6$~$\mu$A~turns$/\Phi_{0}$. The current in the large winding was approximately $\pm 5$~nA, giving a full-signal ampere-turns product of $100$~$\mu$A~turns, corresponding to a flux of $16.7$~$\Phi_{0}$. A flux of $1$~$\mu \Phi_{0}$ therefore corresponds to $0.06$~ppm of the full signal in the ULCA $G_{\text{I}}$ calibration, three times smaller than the observed discrepancy in the electron pump current. However, no investigations have yet been carried out on the performance of our CCC in the low-flux regime, so the size of possible noise-rectification errors is not known. Low flux ratio accuracy tests such as those presented in Ref. \onlinecite{drung2015improving} should provide useful information on the scale of possible errors. We note that if these errors are affecting the ULCA calibrations in our experiment, they are remarkably constant in time, as shown by the $\sim 5 \times 10^{-8}$ relative stability of the ULCA input gain over the duration of the measurement campaign illustrated in supplementary section E. This indicates that if noise is affecting the SQUID, its most likely source is the CCC bridge electronics, rather than external sources.

The upper frequency limit for accurate pumping with tunable-barrier pumps has previously been empirically established at around $1$~GHz \cite{giblin2019evidence}. We have shown that this can be increased, albeit in a rather exceptional sample of pump. In this study, the practical upper frequency limit was determined by a combination of plateau rounding, and increased incidence of switching events which shifted the pump operating point in the $V_{\text{ENT}}-V_{\text{EXIT}}$ plane. This hints at device-physics factors which may limit the practical upper operation frequency, possibly charge traps which are activated by high frequency components present in the drive signal. Further investigation of more samples of pump could shed fruitful light on this question.

\section{conclusions}

Precision measurements have been made of the current from a silicon electron pump driven at a frequency of $2$~GHz using a custom drive waveform applied to the entrance gate. The pump current is invariant in exit gate voltage with a precision of $0.1$~ppm (32 aA), but offset by roughly $0.2$~ppm from the expected current corresponding to one electron for each pump cycle. The application of a blind measurement protocol provides added confidence that this result is not affected by experimenter bias. At this accuracy level, the measurement of the pump current challenges the state of the art in existing electrical metrology methods, with scaling of small currents using CCCs at low flux levels posing a particularly interesting problem. The recent demonstration of current plateaus due to the dual Josephson effect \cite{shaikhaidarov2022quantized} raises the possibility of a metrological investigation of the dual Josephson effect in the near future, providing added motivation for a better understanding of low current scaling.

\begin{acknowledgments}
The authors would like to thank Colin Porter and Scott Wilkins for making the NPL primary Josephson voltage standard available, and for assistance with setting up the voltmeter calibration. This research was supported by the UK department for Business, Energy and Industrial Strategy. A.F. and G.Y. are supported by JSPS KAKENHI Grant Number JP18H05258.
\end{acknowledgments}

\bibliography{SPGrefs_BlindTestPaper}

\newpage
\section{Supplementary information}

\setcounter{figure}{0}
\renewcommand{\thefigure}{S\arabic{figure}}

\subsection{AWG waveform at 1 GHz}

\begin{figure}
\includegraphics[width=9cm]{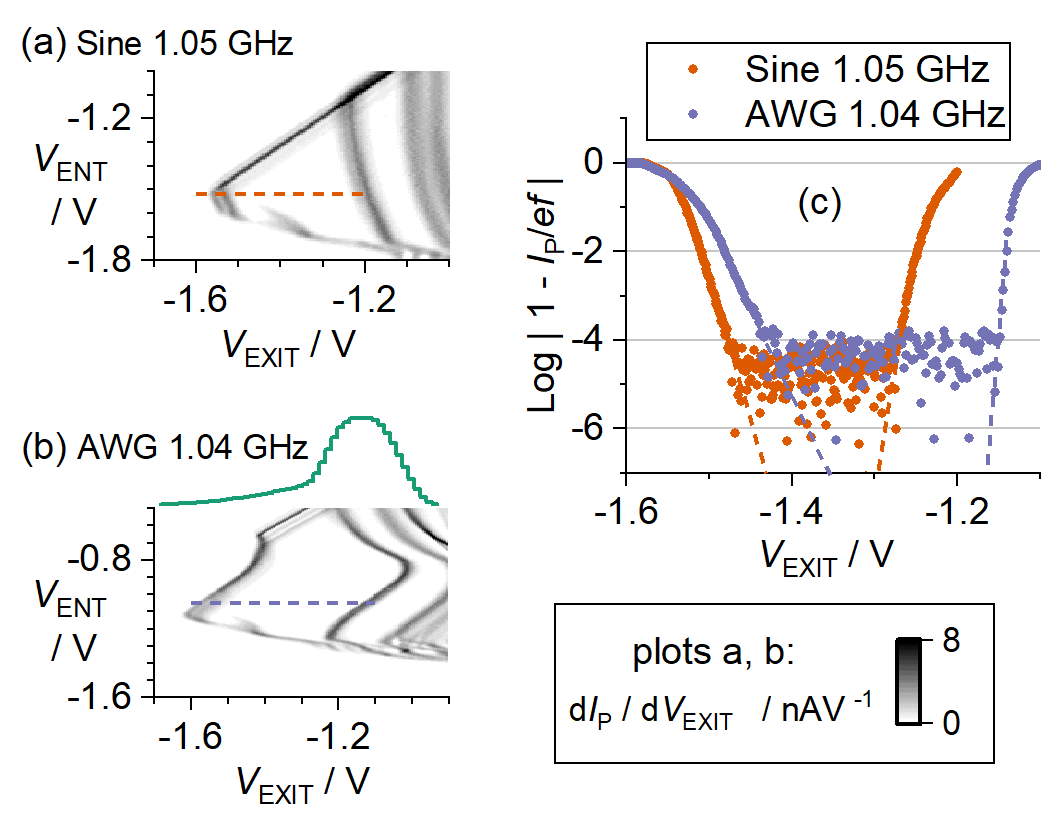}
\caption{\label{AWGSineFig}\textsf{(a): Grey-scale derivative pump map using sine wave drive at 1.05 GHz, $P_{\text{RF}}=11.6$~dBm. (b): Pump map using a waveform from an AWG at repetition rate $f=1.04$~GHz. One cycle of the AWG waveform is shown in the inset. Note that this waveform is subsequently amplified by an inverting amplifier to yield the correct polarity of gate voltage, whereby the negative voltage pulse on the entrance gate raises the entrance barrier to pump an electron. (c): Log-scale plots of the pump current along the horizontal dashed lines in plots (a) and (b).}}
\end{figure}

The silicon pump in this study has already exhibited robust quantisation at $1.05$~GHz with sine wave drive\cite{giblin2020realisation}, as illustrated in the pump map and log plot of figure \ref{AWGSineFig} (a) and (c). As an initial part of the setup process, we tested the pump operation using an AWG waveform at a similar frequency of $1.04$~GHz. This resulted in a substantially wider plateau, seen by comparing the log plots with sine wave and AWG drive in figure \ref{AWGSineFig} (c). Note that the AWG waveform leads to substantial distortion of the pump map (figure \ref{AWGSineFig} (b)), due to the electron capture occurring at different rates $dV_{\text{ENT}}/dt$ as $V_{\text{ENT}}$ is scanned. This data was an important motivator towards the main study because it showed for the first time that the type of waveform first used on GaAs pumps in Ref. \onlinecite{giblin2012towards} could also yield a substantial improvement in plateau flatness with Si pumps.

\subsection{Exploration of higher frequencies}

\begin{figure}
\includegraphics[width=9cm]{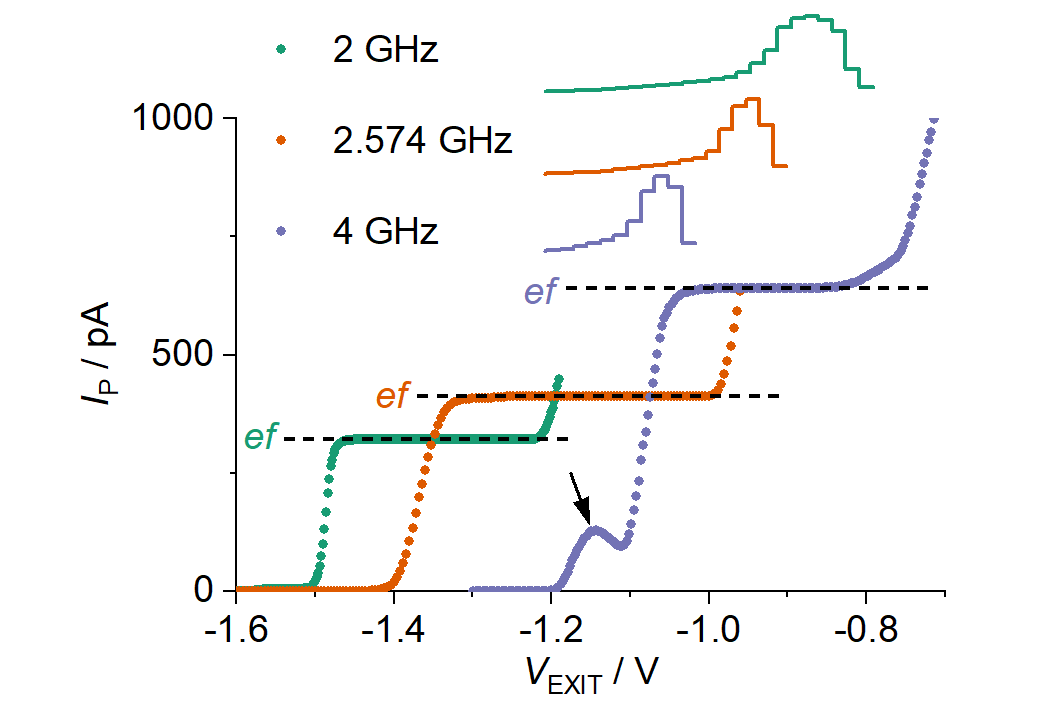}
\caption{\label{HighFreqFig}\textsf{Exit gate scans of pump current using AWG waveforms. The waveforms are illustrated as insets at the top of the plot on a common time axis. Dashed horizontal lines indicate the current $ef$ at each frequency. Entrance gate voltages are $-0.79$~V, $-1.08$~V, and $-1.35$~V at frequencies of $2$~GHz, $2.574$~GHz and $4$~GHz respectively. The AWG output amplitude was 0.47 V pp for all measurements prior to amplification by a $15$~dB wide-band inverting amplifier. The arrow indicates a feature possibly due to non-adiabatic excitation.}}
\end{figure}

\begin{figure*}
\includegraphics[width=18cm]{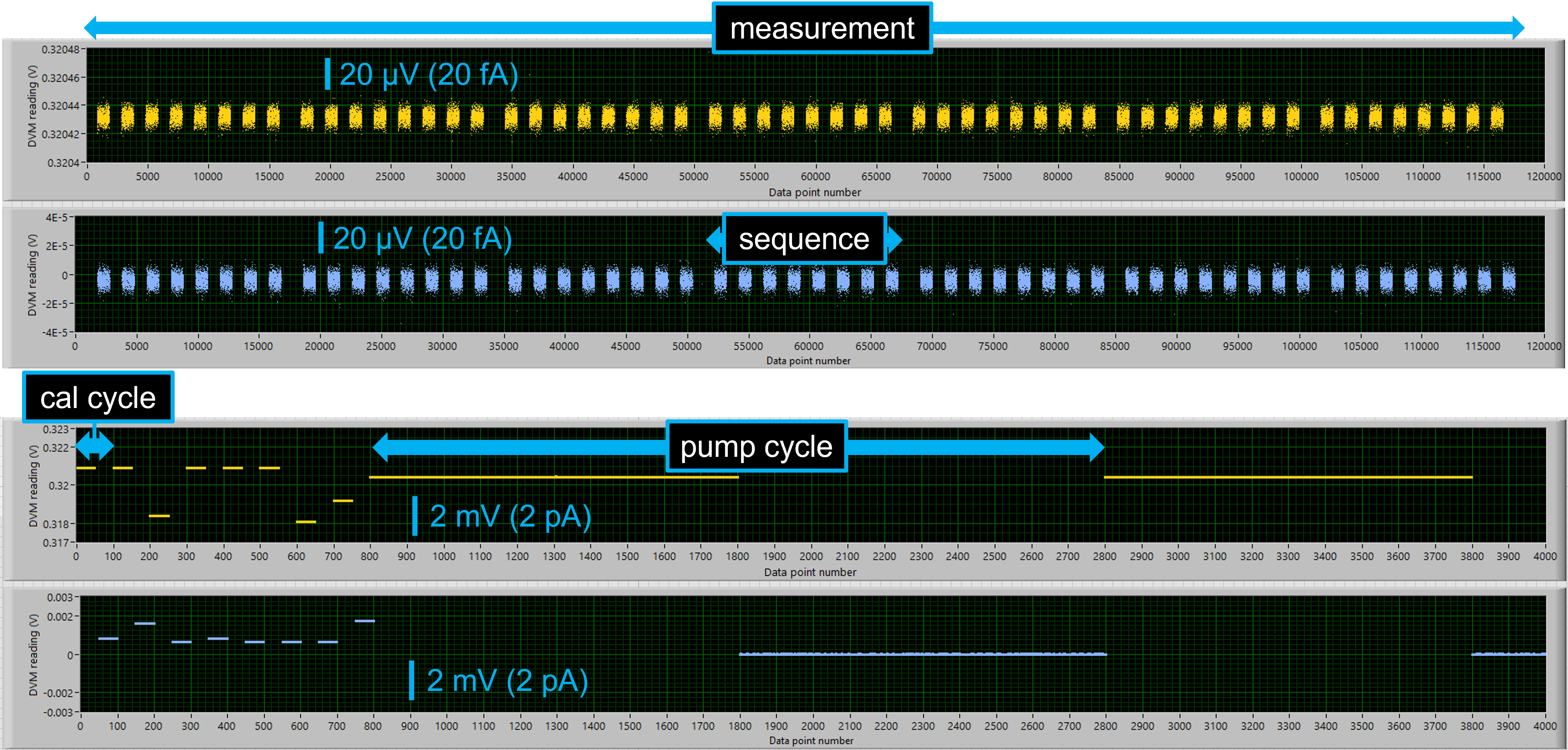}
\caption{\label{RawDataFig}\textsf{Raw data, as viewed in a LabView program used to visualise the data during the measurement campaign. The top pair of plots show the raw voltmeter data from one measurement - run 16, measurement 3. The plots are zoomed to highlight the `on' (upper plot, yellow points) and `off' (lower plot, blue points) pump data. The voltmeter calibration data are off the scale of these plots. The lower pair of plots show the beginning of the measurement on an expanded y-axis, and a zoomed x-axis. The first 800 data points are voltmeter calibrations. The x-axis is simply the sequential data point number. This does not quite map linearly onto time, because the cal cycles used a voltmeter auto zero with every data point, whereas an auto zero was performed after every 25th data point for the measure cycles. All data points were integrated over 10 power line cycles. On both plots, vertical bars indicate the y-scale in raw voltmeter units (ULCA input current).}}
\end{figure*}

During the setup of the experiments reported in the main text, frequencies above $2$~GHz were explored using custom waveforms (figure \ref{HighFreqFig}). The data at $4$~GHz shows a feature which may be attributable to non-adiabatic excitation \cite{kataoka2011tunable} resulting from the rapid deformation of the confining potential formed by the entrance and exit gates. Although the $1ef$ plateau at $4$~GHz looks superficially flat on this expanded current scale, its slope could easily be resolved by zooming the data and no precision measurements were attempted. The plateau at $2.574$~GHz was sufficiently flat for metrological investigation, but the stability of the pump map was degraded compared to $2$~GHz, with sudden shifts along the entrance and exit gate axes becoming common on time-scales of a few hours. Switches in the pump state generally occurred more frequently as $f$ was increased, and we speculate that high frequency components in the drive signal may activate charge traps in the device structure. Consequently, all the precision measurements reported in the main text used $f=2$~GHz, with the waveform shown in the inset of figure \ref{HighFreqFig}, and also the inset of figure 1 (b) of the main text.

\subsection{Raw data}

The measurement apparatus and procedure, with two exceptions, are the same as described in Ref. \onlinecite{giblin2020realisation} and its supplementary information. The exceptions are firstly, the use of a blind protocol as discussed in the main text, and secondly, the use of a noise-optimised ULCA \cite{krause2019noise} instead of a standard ULCA \cite{drung2015ultrastable}. All measurements are performed as on-off cycles. For pump measurements, the `on' and `off' states correspond to the entrance gate drive waveform from the arbitrary waveform generator (AWG) being turned on and off respectively. For calibrations of the digital voltmeter (DVM) used to read out the ULCA, the `on' and `off' states correspond to the Josephson voltage standard programmed to output $0.32$~V and $0$~V respectively. 

\begin{figure}
\includegraphics[width=9cm]{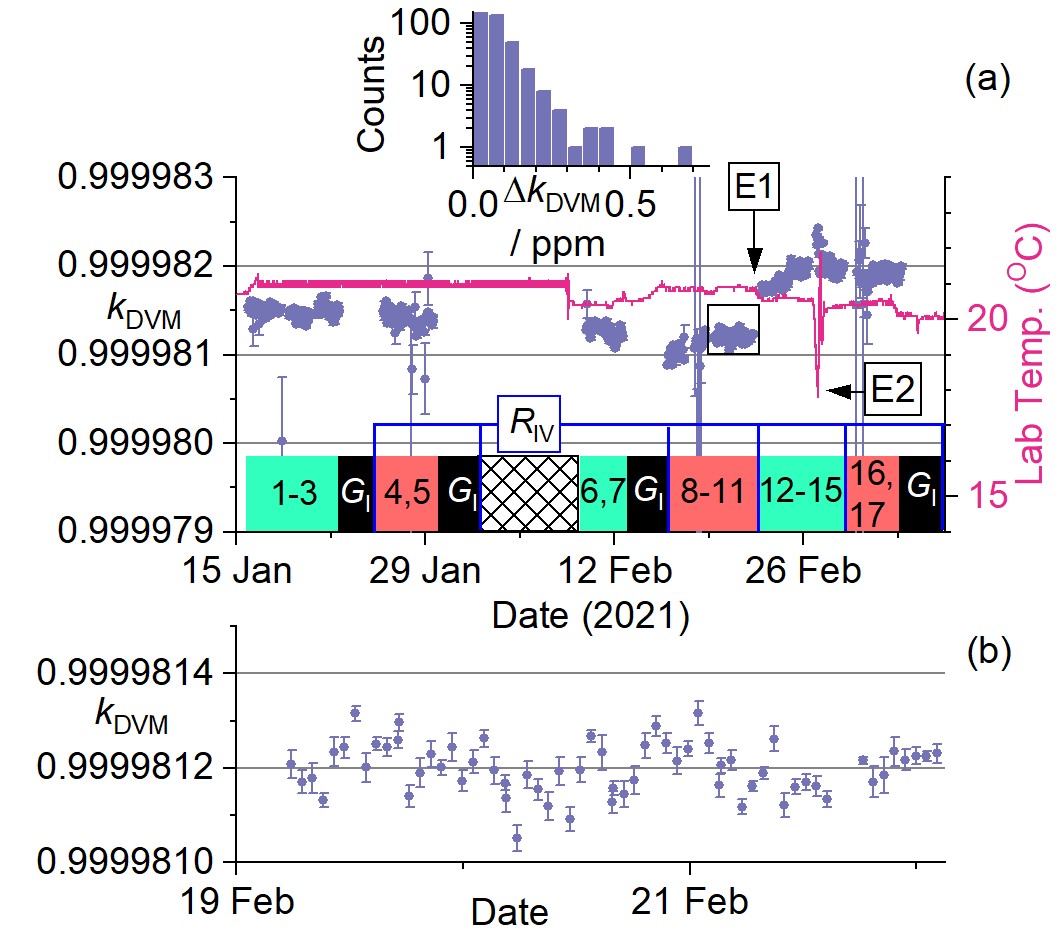}
\caption{\label{TimeLineFig}\textsf{Plot (a): filled circles, left axis: Calibration factor $k_{\text{DVM}}$ of the DVM. line, right axis: laboratory temperature. Coloured blocks at the bottom of the plot, and events labeled `E1' and `E2' are explained in the supplementary text. Inset: Log-scale histogram of the difference between adjacent measurements of $k_{\text{DVM}}$. The black square in the main plot shows a range of DVM calibration data plotted on expanded axes in plot (b).}}
\end{figure}

In figure \ref{RawDataFig} we illustrate some raw data, and explain the nomenclature used to describe the data files. The illustrated data is measurement 3 from run 16. The data are the blind-scaled readings of the Agilent 3458A DVM, connected to either the Josephson voltage standard for the calibration cycles, or the ULCA for the measure cycles. For the calibration cycles, the data are the completely raw readings from the voltmeter, and for the measure cycles the raw readings have been multiplied by the blind scaling factor $\beta=1.00000387$. The particular measurement illustrated here consisted of 7 `sequences'. Each sequence starts with 8 voltmeter calibration cycles. The calibration cycles were done with the DVM auto zero turned on, and 50 data points for each on or off segment. After the calibration cycles, the voltmeter was connected to the ULCA output, and a set of pump measurement cycles were done with 1000 data points for each segment, auto zero off, and an auto zero operation every 25 data points (optimisation of the DVM auto zero interval in the context of single-electron pump measurements was first discussed in Ref. \onlinecite{stein2016robustness}). For the illustrated measurement, there were 8 measurement cycles in one sequence. Other measurements in the campaign used from 7 to 11 cycles per sequence. After the 7 cal-measure sequences, a final set of 8 calibration cycles was performed, so that each set of measure cycles had a calibration cycle before and after, for evaluating the calibration factor to apply to the measurement data as described in the supplementary information to Ref. \onlinecite{giblin2020realisation}. The data analysis evaluated the pump current separately for each sequence, and the statistical properties of this data was used as a pass / fail criteria for the measurement, as described in supplementary section G. The current reported for the measurement was the weighted mean over the sequences. 

Two points are worth remarking in the data. The first is that the hysteretic Josephson voltage standard does not always yield the same step number (it was programmed to switch between nominal values of 320 mV and 0 V). As discussed in the supplementary information to Ref. \onlinecite{giblin2020realisation}, this is not an issue as long as the DVM is linear over the narrow range of voltages sampled by the different calibration steps. The second is the remarkable stability of the ULCA offset. By eye, it does not appear to drift by more than about 1 fA over the course of the measurement. We will examine the stability of the ULCA gain and offset in more detail in supplementary section E.
 
\subsection{Voltmeter calibrations and measurement time-line}

In figure S4 we have combined several pieces of information pertinent to the measurement campaign. The main graph of plot (a) shows, on the left axis, the calibration factors, $k_{\text{DVM}}$, of the DVM recorded during the measurement campaign. We define the calibration factor as $k_{\text{DVM}} \Delta V_{\text{IND}}= \Delta V_{\text{REF}}$, where $\Delta V_{\text{IND}}$ is the change in indicated voltage and $\Delta V_{\text{REF}}$ is the change in applied reference voltage evaluated from an on-off cal cycle. Each plotted point is averaged from a set of 8 calibration cycles directly against the Josephson array at a nominal voltage of $0.32$~V. No data points have been omitted from this plot, and some outlying data points with large error bars are the result of failure of the frequency lock to the Josephson array control electronics. The pink line plotted on the right axis shows the laboratory temperature, as measured by a sensor integrated into the ceiling. Periods when the experiment was not running are visible as gaps in the voltmeter calibration data, and to clarify the experimental time-line, shaded blocks at the bottom of the plot indicate what was happening. Four types of activity are indicated: The experimental runs, numbered 1-17; the weekend calibrations of the ULCA input stage gain $G_{\text{I}}$; The short calibrations of the ULCA output stage $R_{\text{IV}}$, and finally a period of down-time indicated by a cross-hatched block when the experiment was stopped due to a fault in the AWG used to generate the pump drive signal. 

Two events marked E1 and E2 are indicated. E1 marks when an un-used instrument in the experimental rack (a sine wave generator) was switched off. The reduction of heat produced in the rack caused a noticeable change in the calibration factor of the voltmeter, which was mounted directly above the sine wave generator. The fact that this is visible in the $k_{\text{DVM}}$ data illustrates the sensitivity of the direct calibrations of the DVM against the Josephson array. The event E1 also lowered the temperature of the ULCA, mounted higher up in the rack, reducing $A_{\text{TR}}$ by roughly $0.15$~ppm. Event E2 marks a dramatic excursion of the laboratory temperature caused by planned maintenance of the air conditioning. This resulted in a larger uncertainty assigned to some of the measurements of run 15 because of rapid changes in the ULCA temperature. The transition from stable to fluctuating temperature roughly half-way through the measurement campaign was co-incident with a transfer of liquid helium into the experimental dewar. It may also be related to increased activity in adjacent laboratories as activities were re-started and staff returned following relaxation of covid-19 control measures.

One important contribution to the uncertainty of the current measurement is the stability of the DVM on the 1-hour time taken for a cal-measure sequence. The inset to figure \ref{TimeLineFig} (a) shows a histogram of the difference in $k_{\text{DVM}}$ between adjacent calibrations during measurements, denoted $\Delta k_{\text{DVM}}$. Generally, the DVM is stable to better than $0.2$~ppm on time-scales of an hour, but jumps in $k_{\text{DVM}}$ of up to $0.5$~ppm sometimes occur. As in our previous study\cite{giblin2020realisation}, the uncertainty due to the drift in $k_{\text{DVM}}$ was evaluated using a rectangular distribution as $\Delta k_{\text{DVM}} / 2 \sqrt{3}$, so a jump in $k_{\text{DVM}}$ of $0.2$~ppm contributes $0.057$~ppm to the combined uncertainty in the pump current. The 1-hour DVM calibration interval is therefore consistent with achieving a combined uncertainty in the pump measurement of $0.1$~ppm. To visualise the short-term stability of the DVM in the time domain, plot (b) shows a portion of the main plot on an expanded time axis. Over this 3-day period, the voltmeter calibration did not drift by more than $0.3$~ppm. The voltmeter calibration data are of general interest for electrical metrology, where voltmeters such as the 3458A are commonly used as transfer standards. From the general perspective of evaluating the DVM performance in metrological applications, this data set shows the DVM comfortably exceeding its manufacturer's 24-hour accuracy specification of $1.5$~ppm on the 1 V range. Calibrations over longer time-scales (not shown) show that the 90-day specification of $4.6$~ppm is also exceeded by typically a factor $5$.

\subsection{ULCA calibrations}

\begin{figure}
\includegraphics[width=9cm]{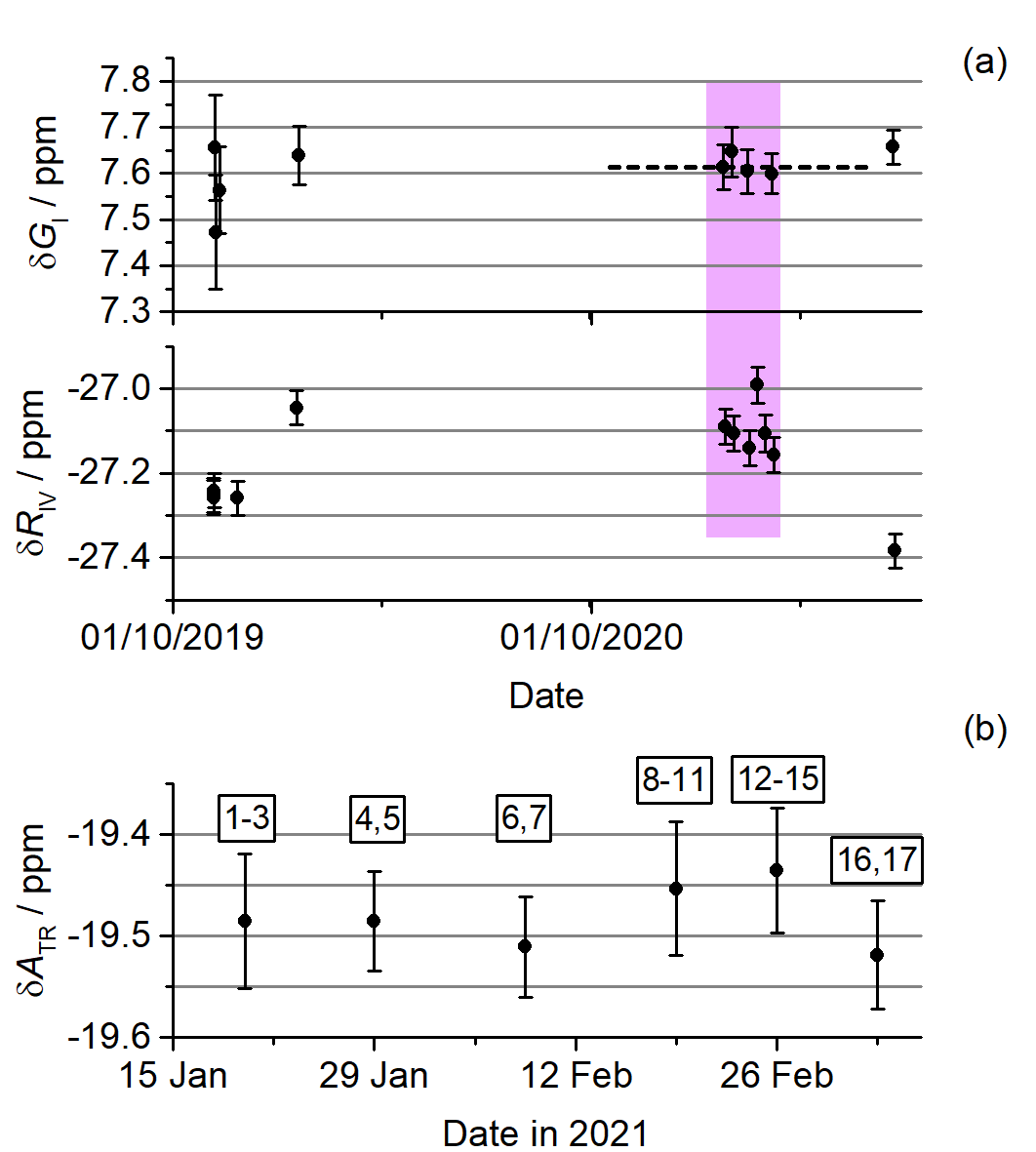}
\caption{\label{ULCACals}\textsf{(a): Deviations from nominal of (upper plot): ULCA input current gain and (lower plot): ULCA output stage gain, corrected to a standard temperature. The plot shows all the calibrations performed on this ULCA since its delivery to NPL. The time period covered by the measurement campaign is shown as a purple shaded box, and the fixed value adopted for the input stage gain during the measurement campaign is shown as a horizontal dashed line. (b): Deviation from nominal of the ULCA transresistance gain calculated from the data in plot (a), for use during the measurement campaign. The boxes above each data point show the run numbers covered by the 6 values of trans-resistance gain.}}
\end{figure}

\begin{figure}
\includegraphics[width=9cm]{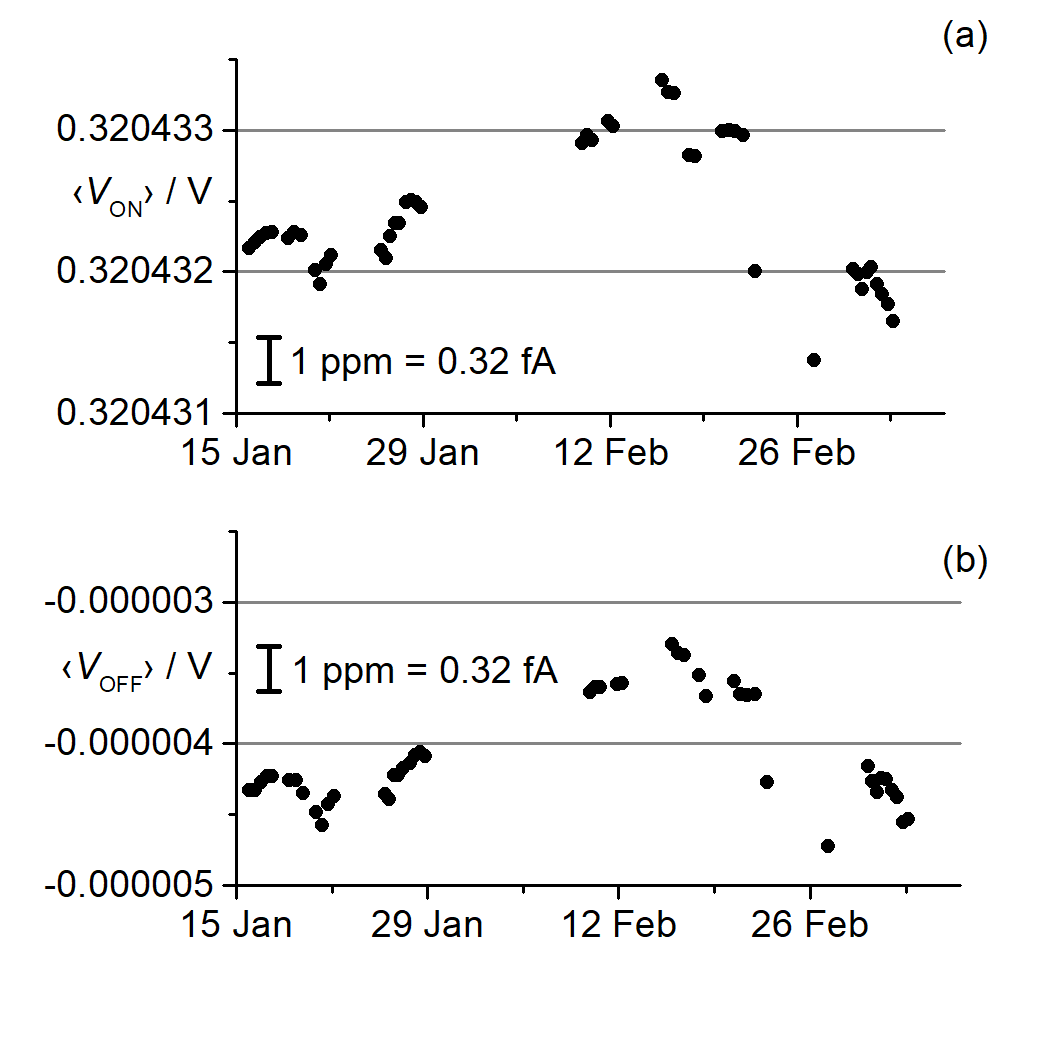}
\caption{\label{ULCAOffset}\textsf{Averaged DVM readings with the pump (a): on, and (b): off. Each data point is averaged from one measurement, so is the mean of $\sim 60,000$ DVM readings. A scale bar indicates 1 ppm of the $320$~pA pump current.}}
\end{figure}

The noise-optimised ULCA was calibrated using a cryogenic current comparator (CCC) bridge, as described in Ref. \onlinecite{giblin2019interlaboratory}. For the calibrations, the ULCA was hand-carried to an adjacent laboratory. It was specifically carried by hand rather than on a trolley to minimise the possibility of mechanical shocks. As illustrated in the time-line of figure \ref{TimeLineFig} (a), a total of 4 calibrations of the input stage gain $G_{\text{I}}$ (nominal value 1000), and 6 calibrations of the output gain $R_{\text{IV}}$ (nominal value $1$~M$\Omega$) were preformed during the measurement campaign. The overall trans-resistance gain of the ULCA is $A_{\text{TR}} = G_{\text{I}}R_{\text{IV}}$ (nominal value $1$~G$\Omega$) \cite{drung2015ultrastable}. The results of all calibrations of this ULCA unit since its delivery to NPL are shown in figure \ref{ULCACals} (a). The historical behaviour of the input and output gains is different, and resulted in different statistical treatments. The input stage gain does not show any significant drift over the measurement campaign, and furthermore, the  limited number of additional calibrations before and after the campaign did not give any evidence for long-term drift. Consequently it was assumed to be constant during the measurement campaign. Its value was taken to be the weighted mean of the four calibrations during the campaign, shown as a horizontal dashed line in figure \ref{ULCACals} (a). On the other hand, the output stage gain shows some drift over time. Values of $R_{\text{IV}}$ were chosen half way between `before' and `after' calibration values, with uncertainties which included a drift term derived from a rectangular distribution. In this way, five values of $A_{\text{TR}}$ were calculated to cover runs 4-17. Runs 1-3 were not preceded immediately by any ULCA calibrations, so the value of $R_{\text{IV}}$ was taken to be the first $R_{\text{IV}}$ calibration, in between runs 3 and 4, with an uncertainty derived from a rectangular distribution bounded by the highest and lowest $R_{\text{IV}}$ calibrations during the measurement campaign. In other words, we assumed that the drift behaviour of $R_{\text{IV}}$ for the few days covering runs 1-3 was similar to the behaviour during the rest of the measurement runs. The 6 values of $A_{\text{TR}}$ with their combined standard uncertainties used to analyse the measurements are shown in figure \ref{ULCACals} (b). 

The remarkable stability of the ULCA offset current is already visible in the raw data of figure \ref{RawDataFig} (a), and in figure \ref{ULCAOffset} we go further and show the averaged values of the `ON' and `OFF' signals measured by the DVM. Each data point in this graph is the average of all the ON (plot (a)) or OFF (plot (b)) DVM readings after rejecting the first 300 readings in each segment. The offset current does not change by more than $2$~fA over the 2-month period covered by the measurements. The drift in offset current may be partially attributable to changes in ULCA temperature, but there may also be contributions due to changes in leakage currents through the electron pump control gates. The possible leakage current paths through the device gates were discussed in the supplementary information to Ref. \onlinecite{giblin2020realisation}.

\subsection{ULCA linearity}

\begin{figure}
\includegraphics[width=9cm]{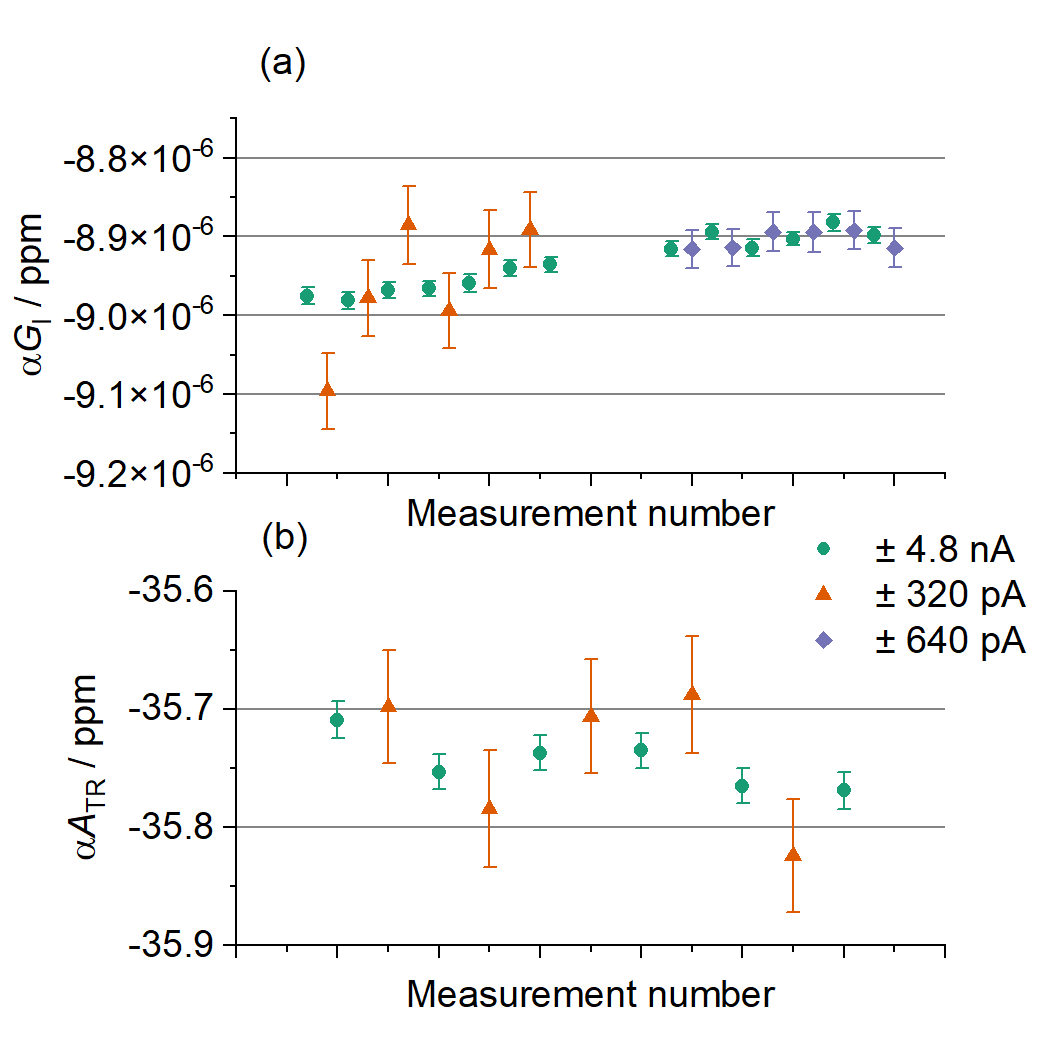}
\caption{\label{LinGraphFig}\textsf{(a): Difference in input current gains $G_{\text{I}}$ of two ULCA units for two series of measurements in self-test configuration, in which the test current was alternated between a `high' current of $4.8$~nA and a `Low' current, either $320$~pA or $640$~pA. (b): Difference in trans-resistance gains $A_{\text{TR}}$, alternating the test current between $4.8$~nA and $320$~pA. The data plot legend refers to both panels (a) and (b).}}
\end{figure}

The linearity of the ULCA gain is a key assumption in this experiment, because the calibration of $G_{\text{I}}$ is done at an input current of $\sim 5$~nA and the pump current during the measurement is $320$~pA. One previous investigation set an upper bound on the non-linearity of the overall UCLA transresistance gain $A_{\text{TR}}$ at around the $0.1$~ppm level \cite{krause2019noise}. We attempted to reduce this upper bound, using two test methods previously demonstrated for the ULCA. First, we compared the input stage current gains of two ULCA units, as was first demonstrated in Ref. \onlinecite{drung2015ultrastable}. This is called the `self-test' configuration. A standard ULCA unit, not otherwise used in our experiment, was used as a source to generate a test current for comparing its input stage gain $G_{\text{I,source}}$ with the input stage gain of the noise-optimised experimental ULCA $G_{\text{I,measure}}$. This self-test configuration is quite straightforward to implement, because the readout DVM measures a small signal derived from the difference in the input gains of the two ULCAs, denoted $\alpha G_{\text{I}} = G_{\text{I,source}} - G_{\text{I,measure}}$. We alternated sets of forward-reverse cycles with test currents of $\pm 4.8$~nA, $\pm 320$~pA and $\pm 640$~pA to obtain the data of figure \ref{LinGraphFig} (a). The forward-reverse cycle time was $60$~s, and the data points are averaged from 100 and 1000 cycles for the $\pm 4.8$~nA and $\pm 320$~pA currents respectively. The background drift of $\alpha G_{\text{I}}$ visible in the high current data is due to temperature variation of the ULCAs, but by evaluating the difference between each low-current data points (orange triangles) and the mean of the two adjacent high current data points (green circles), we can extract the current dependence as a mean over 6 cycles of high-low-high current. We obtain the current dependence in $\alpha G_{\text{I}}$ between $4.8$~nA and $320$~pA as $0.002 \pm 0.029$~ppm. An additional run examined the current dependence between $4.8$~nA and $640$~pA (blue diamonds). This data was not evaluated, but clearly the current dependence is around a part in $10^8$ or less. 

For the second test, we measured the current dependence of the difference in the overall trans-resistance gains of the two ULCAs, again with the standard ULCA in `source' mode, and the noise-optimised experimental ULCA in `measure' mode. This test configuration is illustrated in figure 6 of Ref. \onlinecite{krause2019noise}. It is less straightforward to implement than the self-test configuration, because the voltage outputs of the source and measure ULCAs have opposite signs. We implemented a protocol equivalent to figure 7b of Ref. \onlinecite{krause2019noise}. A single DVM could be connected to either the source or measure ULCA using an automated switch - the same switch that was used in the main experiment to connect the DVM either to the ULCA output or the JVS. One cycle consisted of four segments of data: the test current was applied with both polarities with the DVM connected to the source ULCA, recording a forward-reverse difference voltage $\Delta V_{\text{source}}$ and then the test current was applied with both polarities with the DVM connected to the measure ULCA, recording a difference voltage $\Delta V_{\text{measure}}$. Acquiring one cycle took 2 minutes. Assuming that the DVM calibration factor does not change on this time-scale, The ratio of ULCA transresistance gains is given by $\alpha A_{\text{TR}} = A_{\text{TR,source}} / A_{\text{TR,measure}} = \Delta V_{\text{measure}} / \Delta V_{\text{source}}$. We are interested in whether the ratio of gains depends on current, so as in the tests of $G_{\text{I}}$ linearity, we alternated 1000 cycles at $\pm 320$~pA test current, with 100 cycles at $\pm 4.8$~nA test current to yield the averaged data points in figure \ref{LinGraphFig} (b). Similarly to the data of figure \ref{LinGraphFig} (a), we averaged the high-low-high differences, to obtain the current dependence of $\alpha A_{\text{TR}}$ as $0.006 \pm 0.023$~ppm.

Of course, this data does not conclusively rule out non-linearity in the ULCA unit used for the measurements. It only gives information on the linearity of the difference in the gains of the two ULCA units. It is a slightly stronger test than the one published in Ref. \onlinecite{krause2019noise}, however. While that measurement used two nominally identical noise-optimised ULCAs, our measurement used a standard ULCA in the `source' role. The different values of resistors used in the current scaling networks make it less likely that both ULCA units would have the same current-dependence to the gain.

\subsection{Statistical tests and data set rejection}

\begin{figure}
\includegraphics[width=9cm]{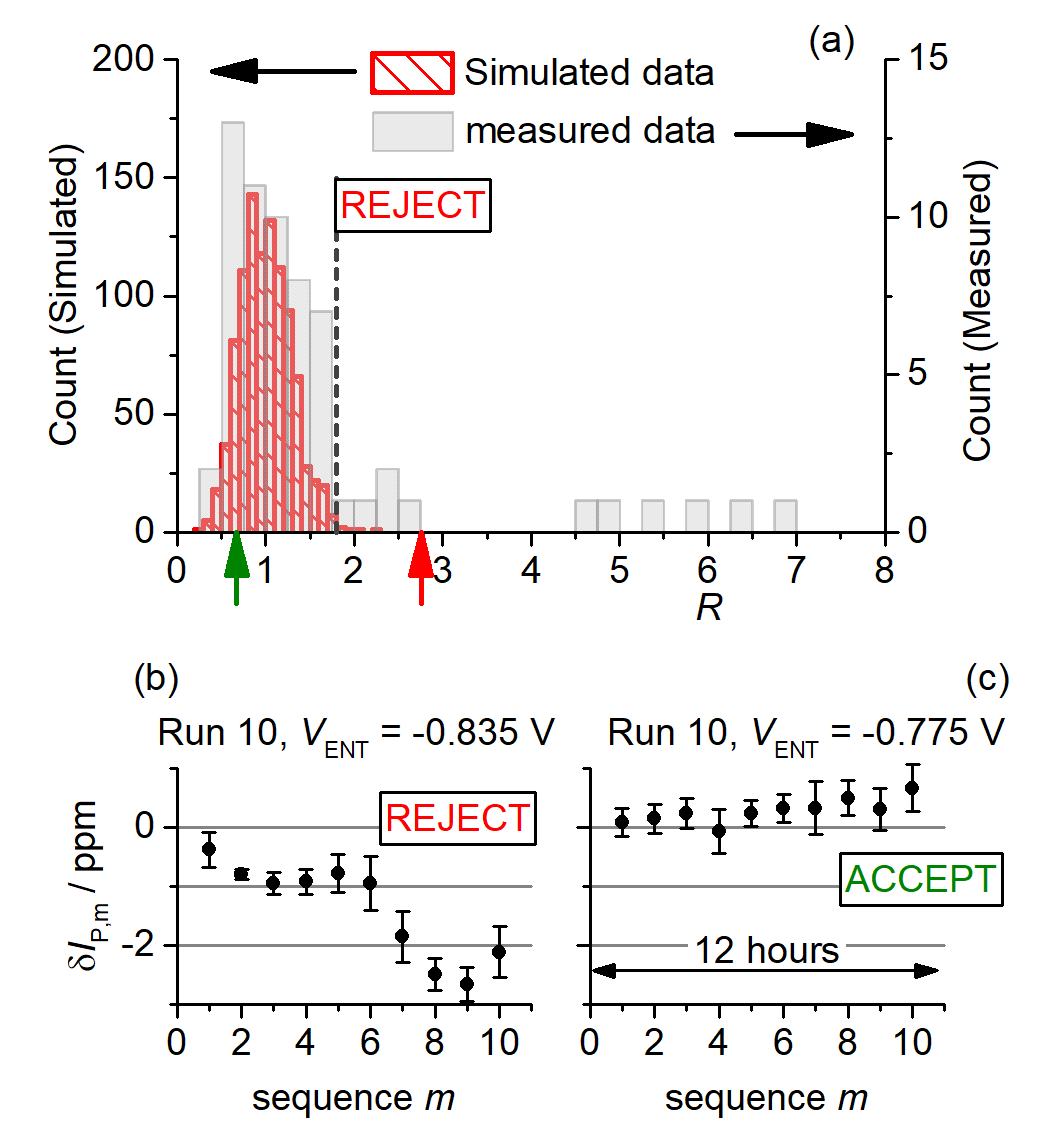}
\caption{\label{StatsGraphFig}\textsf{(a): Histograms of the quantity R, defined in the supplementary text as the ratio of the standard deviation of the $m$ values of $I_{\text{P}}$ calculated for each measurement ($m$ is sequence number) to the average uncertainty of $I_{\text{P,m}}$. The grey bars referred to the right axis are for the measured data, and the red cross-hatched bars referred to the left axis are for $1000$ simulated measurements assuming a stationary mean. A vertical dashed line shows the $R=1.7$ rejection threshold derived from the probability of the simulated data having an $R$ greater than this value. Panels (b) and (c) show $I_{\text{P,m}}$ for two example measurements from run 10, with (b): $R>1.7$ and (c): $R<1.7$. The $R$ values for these data sets are indicated with red and green arrows respectively on panel (a).}}
\end{figure}

As mentioned in the main text, the pump state, as documented by the `pump maps', changed during the measurement campaign, with some obvious dramatic changes occurring during some measurements, and more subtle changes during other measurements. Even if the pump map was stable, some of the measurements close to the edges of the current plateaus could be affected by small fluctuations in offset charge, leading to relatively large changes in pump current as the operating point drifted on and off the plateau. It could not generally be assumed that the pump current sampled by a measurement lasting more than 10 hours represented a stationary mean. Each measurement was therefore subjected to a statistical test. Recall from supplementary section S3, that the pump current from each sequence was evaluated separately. This yielded $m$ values of $I_{\text{P}}$, denoted $I_{\text{P,m}}$ with uncertainties $U (I_{\text{P,m}})$, where $m$ is the number of sequences in the measurement. If all the $I_{\text{P,m}}$ are sampling the same value of pump current, on average the standard deviation of the $I_{\text{P,m}}$, $\sigma (I_{\text{P,m}})$ will be equal to the mean of the uncertainties, $\langle U (I_{\text{P,m}}) \rangle$. We propose the ratio $\sigma (I_{\text{P,m}}) / \langle U (I_{\text{P,m}}) \rangle = R$ as a statistical measure of the stationarity of the data, and in figure \ref{StatsGraphFig}, we plot a histogram of this quantity (grey bars, right axis) for the 64 measurements performed during our campaign. We also plot (red bars, left axis) a histogram of the same quantity obtained from 1000 simulated measurements, in which the simulated raw data, both for the measurement and calibration cycles, was generated from a stationary mean multiplied by Gaussian white noise with the same standard deviation as the real data. As expected, the most probable value of $R$ for this simulated stationary data is $1$, and the probability of obtaining a measurement with $R>2$ from a set of $1000$ measurements becomes negligible. Since we only performed $64$ measurements, we assigned a cutoff of $R=1.7$, and rejected measurements with $R>1.7$. Comparing the histogram of the measured data with the simulation, it is clear that a significant number of data sets have an $R$ value which would be improbably high if the pump current was constant during the measurement. This is actually expected, for the reason that some of the precision measurements were selected with control parameter values close to the edges of the current plateau. For these measurements, small fluctuations in offset charge during the measurement (equivalent to a drift in the control parameters) would cause the pump current to drift away from $ef$.

To see the accept / reject criteria in action, two example measurements from run 10 are plotted in figures \ref{StatsGraphFig} (b) and (c), with the corresponding $R$ values marked with red and green arrows on the x-axis of panel (a). The data of panel (b) clearly shows a decrease in the pump current, and it would be tempting to reject this data set based just on this time-domain visualisation of $I_{\text{P,m}}$. However, the definition of the $R$ parameter makes this otherwise subjective process more quantitative. Altogether, 14 measurements during the entire measurement campaign had $R>1.7$.

\subsection{full data set}

\begin{figure*}
\includegraphics[width=18cm]{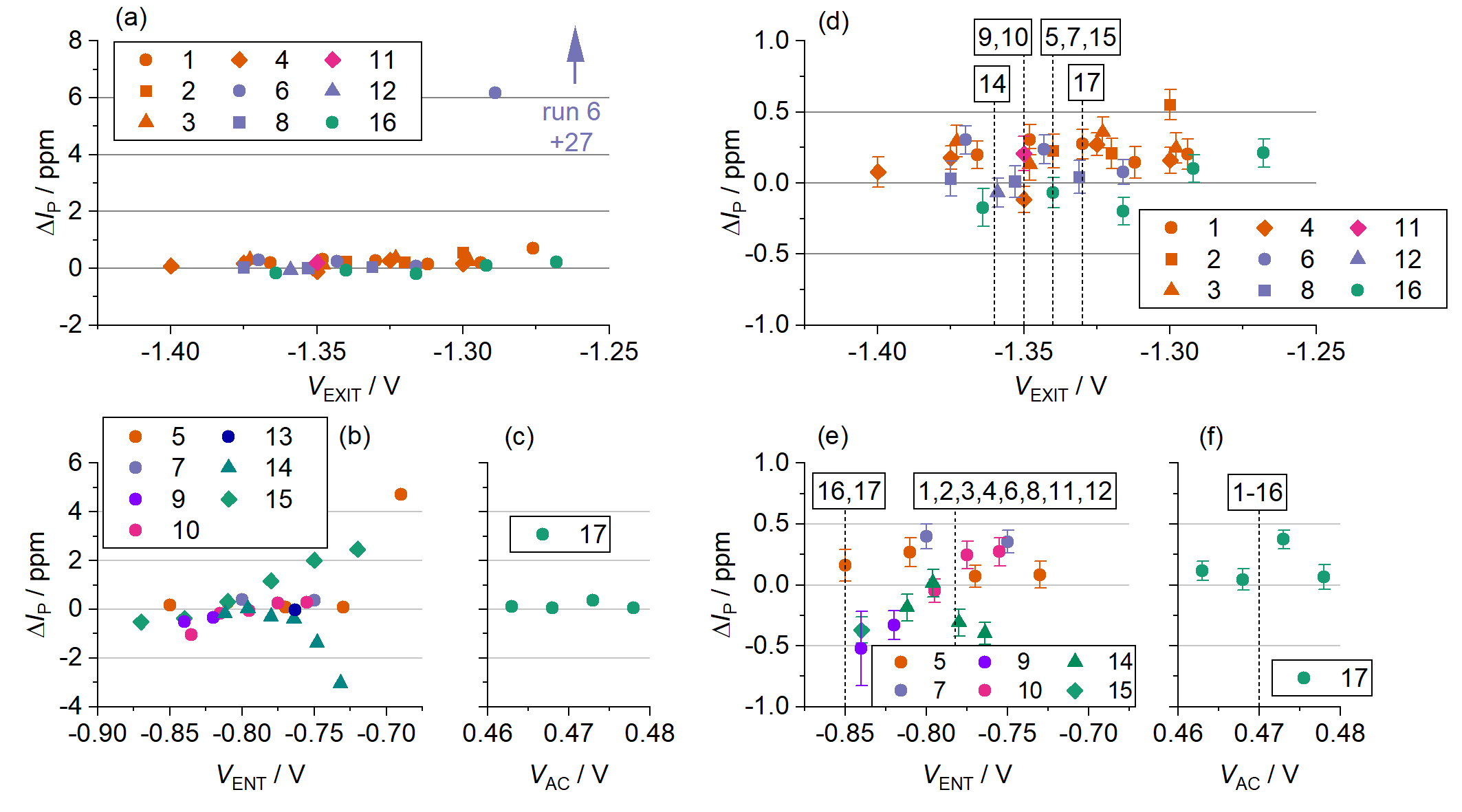}
\caption{\label{PrecisionDataGraph}\textsf{a-c: Deviation of the pump current from its nominal value as a function of (a): Exit gate voltage, (b): Entrance gate voltage and (c): AWG output amplitude. All of the measurements from the 17 runs are shown in these plots. One data point in panel (a) is off the y-axis scale, and is indicated by an arrow. (d), (e) and (f): the sub-set of data in plots (a), (b) and (c) respectively, which passed the stationary mean test, on expanded axes. In each plot, vertical dotted lines indicate fixed values of the scanned parameter for runs in the other plots. Error bars indicate combined standard uncertainties $U_{\text{T}}$.}}
\end{figure*}

\begin{figure*}
\includegraphics[width=17cm]{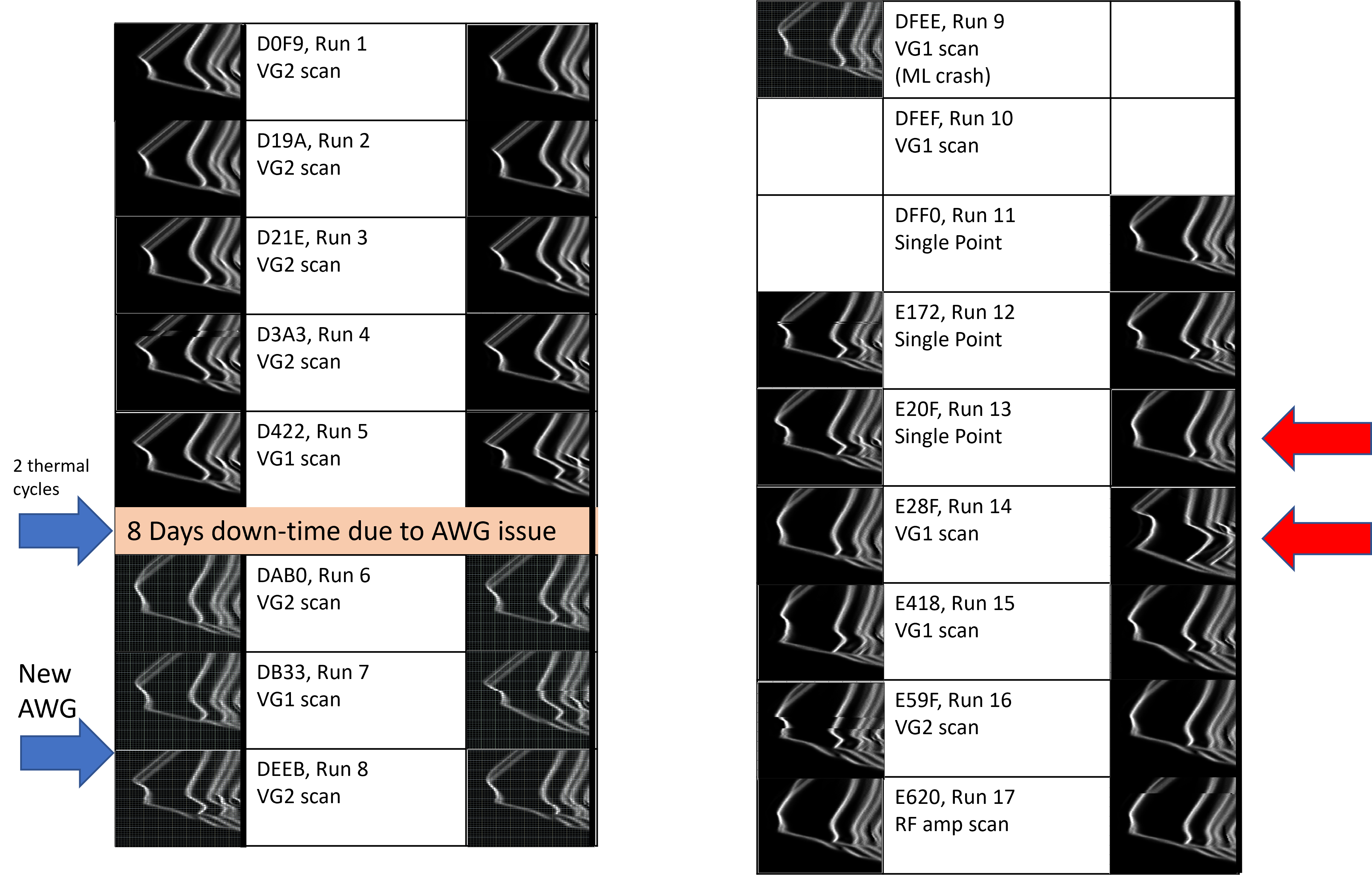}
\caption{\label{PumpMapsFig}\textsf{Thumbnail pump maps measured before and after each precision scan. Each pump map is an inverted grey-scale derivitive plot of the current similar to the one shown in figures 1(b) and 2(a) of the main text. The axis limits are the same for each thumbnail: The x-axis is $V_{\text{EXIT}}$, from $-1.7$~V to $-0.8$~V, and the y-axis is  $V_{\text{ENT}}$, from $-1.4$~V to $-0.4$~V. Each horizontal row of the table represents a precision run. The middle cell contains some text data describing the run, including the 4-digit hexadecimal file number identifying the raw data set for the precision run. The left-most cell shows the pump map recorded before the run, and the right-most cell shows the pump map recorded after the run. Missing pump maps for runs 9,10 and 11 were due to software crashes. Red arrows highlight runs in which the pump map changed dramatically during the run.}}
\end{figure*}

Due to instability of the pump after run 5, only the data from runs 1-5 are analysed in the main text. The increasing instability is visible in the pump maps of figure \ref{PumpMapsFig}, and also in the increasing number of runs which failed the stationary mean test. In figure \ref{PrecisionDataGraph}, we present all of the precision data on linear axes. Plots (a,b,c) show all of the measurements on expanded y-axes, and plots (d,e,f) show the sub-set of the measurements which passed the stationary-mean test. Figure \ref{PumpMapsFig} shows the full set of `fingerprint' pump maps obtained before and after each precision measurement run. For data integrity purposes, this figure also includes the 4-digit hexadecimal file identifier for the precision raw data.

\end{document}